\documentclass[number,preprint,review,12pt]{elsarticle}

\usepackage{graphics}
\usepackage{amssymb}
\biboptions{square}
\journal{Nuclear Instruments and Methods}

\begin{document}

\begin{frontmatter}



\title{A mobile detector for measurements of the atmospheric muon flux in underground sites}

\author[label1]{Bogdan \corref{cor1} Mitrica}
\ead{mitrica@nipne.ro}
\author[label1]{Romul Margineanu}
\author[label1]{Sabin Stoica}
\author[label1]{Mirel Petcu}
\author[label1]{Iliana Brancus}
\author[label2]{Alexandru Jipa}
\author[label2]{Ionel Lazanu}
\author[label2]{Octavian Sima}
\author[label3]{Andreas Haungs}
\author[label3]{Heinigerd Rebel}
\author[label1]{Marian Petre}
\author[label1]{Gabriel Toma}
\author[label1]{Alexandra Saftoiu}
\author[label1]{Denis Stanca}
\author[label1]{Ana Apostu}
\author[label1]{Claudia Gomoiu}

\address[label1]{Horia Hulubei Institute of Physics and Nuclear Engineering (IFIN-HH), Bucharest, Romania, P.O.B.MG-6}
\address[label2]{Department of Physics, University of Bucharest, P.O.B. MG-11, Romania}
\address[label3]{Institut fur Kernphysik, Karlsruhe Institute of Technology - Campus North, 76021 Karlsruhe, Germany}


%\author{}
%\author{Bogdan \corref{cor1} Mitrica}

\cortext[cor1]{Corresponding author}

%\fntext[label3]{trkkk}

%%\author[label1]{Romul Margineanu}
%\ead{mitrica@nipne,ro}
%\cortext[cor1]{Coresponding author}
%\address{Horia Hulubei Institute of Physics and Nuclear Engineering (IFIN-HH)}
%\fntext[label3]{trkkk}

\begin{abstract}
Muons comprise an important contribution of the natural radiation dose in air 
(approx. 30 nSv/h of a total dose rate of 65-130 nSv/h), as well as in underground sites
even when the flux and relative contribution are significantly reduced. 
The flux of the muons observed in underground can be used as an estimator for the depth in mwe 
(meter water equivalent) of the underground site. 
The water equivalent depth is an important information to devise physics experiments 
feasible for a specific site. 
A mobile detector for performing measurements of the muon's flux was developed in IFIN-HH, Bucharest.
Consisting of 2 scintillator plates (approx. 0.9 m$^{2}$) which measure in coincidence, 
the detector is installed on a van which facilitates measurements at different locations at surface 
or underground. The detector was used to determine muon fluxes at different sites in Romania. 
In particular, data were taken and the  values of meter water equivalents were assessed for several locations from
 the 
salt mine from Slanic Prahova, Romania. 
The measurements have been performed in 2 different galleries of the Slanic mine at different depths. 
In order to test the stability of the method, also measurements of the muon flux at surface at 
different elevations were performed.
The results were compared with predictions of Monte-Carlo simulations using the
CORSIKA and MUSIC codes. 
\end{abstract}


\end{frontmatter}

%\linenumbers

\section{Introduction}
\label{Introduction}

The muons in the atmosphere originate from leptonic decays of pions and kaons 
generated by the high-energy collisions which cosmic rays experience with the atoms of the 
Earth's atmosphere.
Muons are unstable particles which decay  into electrons and positrons, respectively, 
with two corresponding neutrinos (electron ($\nu_{e}$) and muon ($\nu_{\mu}$) neutrinos) 
with a life time (in there own reference system) of $\tau_{\mu} =  2.2 \mu s$. 
As leptons, muons are less affected by hadronic interactions and interact weakly with matter.
They penetrate large thicknesses of matter before they are stopped and subsequently decay. 
Hence they are historically known as "penetrating component" of secondary cosmic rays, even 
detectable in deep underground sites.
The cosmic ray muon flux, defined as the number of muons transversing a horizontal element of area per unit of time  \cite{Grieder},
 is of interest for various branches of science, in elementary 
particle physics as "heavy electron", as messenger of astrophysical processes, in environmental and material 
research inducing natural radiation damages, and with a role for cosmogenic production of long living isotopes.
The focus of this paper is on studies of the atmospheric muon flux in the Slanic Prahova underground site. This site is actually under discussion as location of a large detector of the LAGUNA project \cite{Rubbia_1, Autiero, Mosca}. The paper intends to provide some basic information characterising the site, in particular on the underground depth of the salt mine.

A different reason for measuring the flux of the atmospheric muons in underground arises from the practical
necessity of information on the cosmic radiation background for different sites. 
This background does consists not only of muons which have survived the passage through the
rock above, but also of contributions of natural radioactivity and of muon induced radiation like
neutrons, which can play a decisive role for low background experiments \cite{Schreiner}.

For a simple and efficient procedure for measuring the muon flux at different places a mobile device was set-up and operated since autum 2009, registering the muon flux at the surface and in the underground. Determinations of the water equivalent depth
of any underground site could be done in a reasonable time scale. This feature is important in order to establish very
accurately the overburden thickness in water equivalent of matter (mwe). First measurements have been performed
on the underground site of the Slanic-Prahova mine where IFIN-HH operates a low - radiation level
laboratory \cite{}.

\section{The LAGUNA project}
\label{LAGUNA}

LAGUNA \cite{Rubbia_1}  (Design of a pan-European Infrastructure for Large Apparatus studying Grand
Unification and Neutrino Astrophysics) is a research project , supported by the European Union to
setup the infrastructure for a large underground laboratory with a first step to explore adequate
locations for looking for extremely rare events like proton decay or for the experimental research of Dark
Matter.

Seven underground laboratories of Great Britain, France, Spain, Finland, Italy,
Poland and Romania are involved. For LAGUNA, three detector types are considered based on different 
active detection media: MEMPHYS with water
\cite{Memphys}, LENA, a liquid scintillator detector \cite{Lena} and GLACIER using liquid argon \cite{Glacier}. 
The site for LAGUNA experiments will be chosen along different criteria: the depth of
the site i.e. the ability to absorb and shield against high energy muons, the available space and
possibility to install a large volume detector inside (larger than 100.000 m$^{3}$ ), and the natural 
radiation background. The site proposed in Romania is located in the salt mine Slanic Prahova, geographically situated at 45.23$^{\circ}$ N latitude and 25.94$^{\circ}$ E longitude. The elevation of Slanic site is 408 m above sea level at the entrance of the mine. The muon contribution to the natural radiation dose in air has been determined \cite{dose} at 31.3 $\pm$ 0.6 nSv/h at the same elevation and latitude as Slanic site.

\begin{figure}
\begin{center}
\includegraphics[scale=0.6]{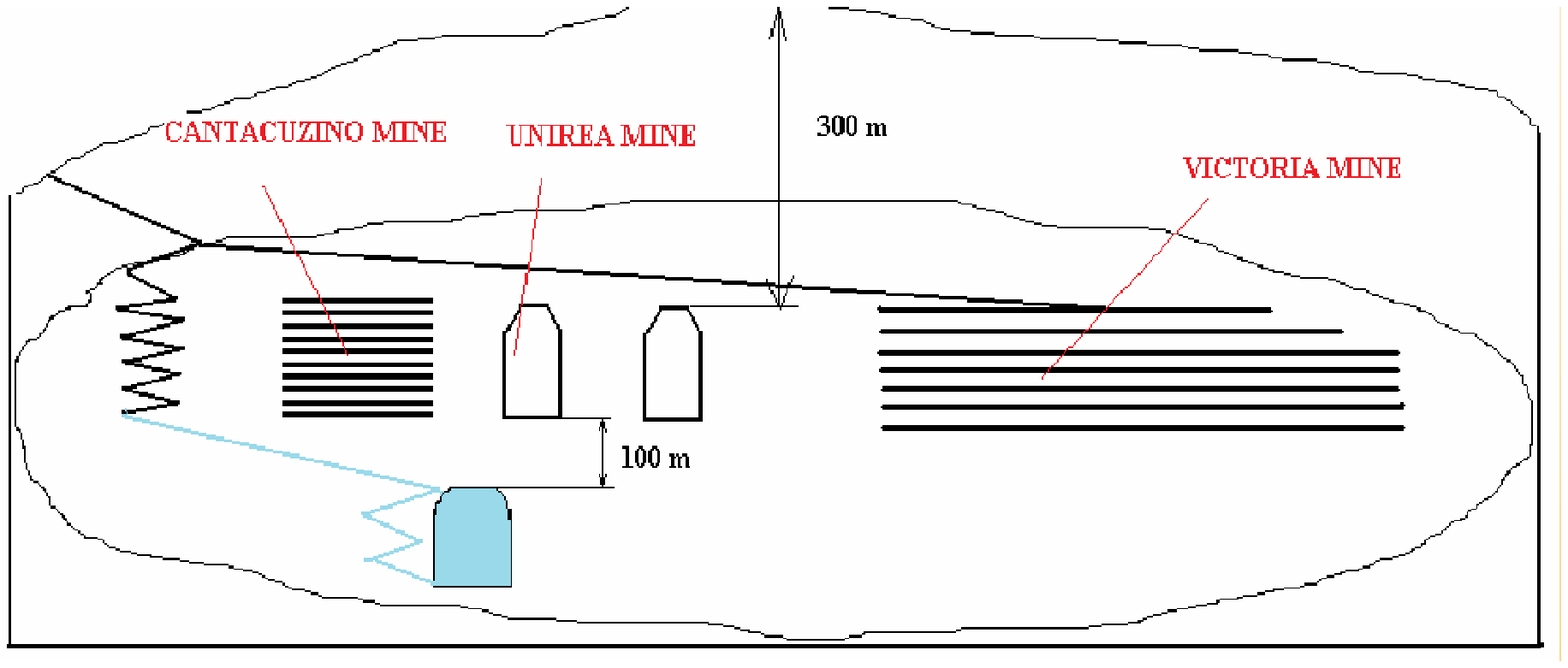}
\caption{\label{fig:slanic} Artistic view of the salt ore of Slanic}
\end{center}
\end{figure}

The salt ore from Slanic consists in a lens of 500 m thickness, few kilometers long and wide 
(see Fig. \ref{fig:slanic}). The salt is extracted from the Slanic mine continuously since ancient 
times and, due to this fact, many galleries (i.e. shaped  caverns)  are already excavated.
 
\begin{figure}
\begin{center}
\includegraphics[scale=0.35]{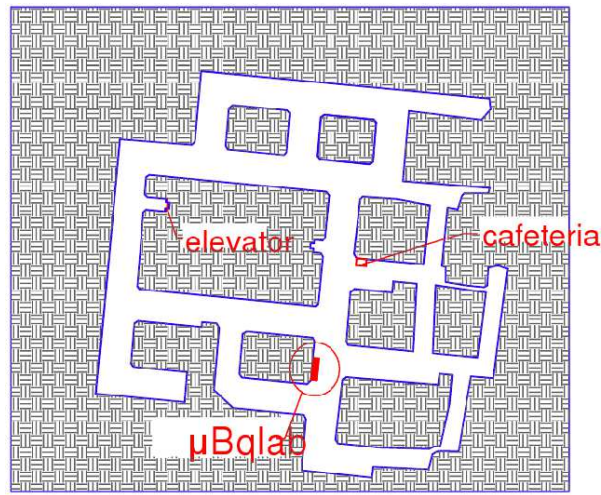}
\caption{\label{fig:unirea} Map of Unirea mine, with the $\mu Bq$ laboratory of IFIN-HH}
\end{center}
\end{figure}

The largest one is the "tourist" mine "Unirea" (see Fig. \ref{fig:unirea}) characterised by:

- depth: 208 m bellow ground level

- temperature: 12.0 -13.0 $^{\circ}$C

- humidity: 65-70 \% 

- excavated volume: 2.9 million m$^{3}$ 

- floor area: 70000 m$^{2}$

- average high: 52-57 m

- distance between walls: 32-36 m

- existing infrastructure: electricity, elevator, phone, Internet, GSM networks.

Besides the "tourist" mine UNIREA,  muon flux measurements have been performed also in the "active" mine CANTACUZINO (see Fig.\ref{fig:slanic}) at two different levels with the physical depths of 188 and 210 m from the entrances.

During the last 5 years a new laboratory  \cite{Margineanu} for low background measurements was 
installed by IFIN-HH in the UNIREA salt mine of Slanic Prahova. From the Slanic site  a huge volume 
of material has been already excavated, but the shallow depth could induce a problem. 
Following  \cite{Rubbia} for the  GLACIER experiment the  Slanic mine could be a feasible location as for this technique
a depth of only 600 mwe is necessary. 
The main goal of this work was to determine the water equivalent depth (mwe) of the Unirea mine.

\section{Monte-Carlo simulation of the muon flux}
\label{Simulations}

Monte-Carlo simulations were used to perform some preliminary explorations regarding the expected results of the  
experimental studies. Different simulation codes have been used:

- CORSIKA 6.735 \cite{CORSIKA} (COsmic Ray SImulation for KAscade), a sophisticated Monte-Carlo code for 
simulations of the development of extensive air showers (EAS) in the atmosphere, has been used 
to estimate the muon flux at surface. As input for the simulations the primary cosmic ray spectrum ( for different primary masses ) in addition to a model for the hadronic interaction of the primaries with the Earth´s atmosphere is necessary (see below).
 
\begin{figure}
\begin{center}
\includegraphics[scale=0.9]{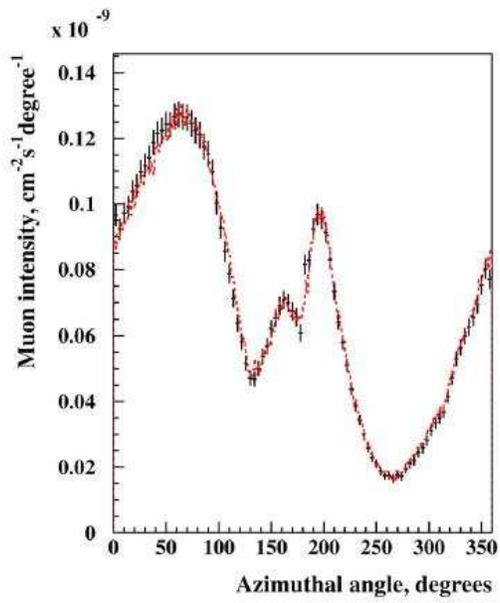}
\caption{\label{fig:LVD} Azimuthal distribution of single muon intensities in the underground Gran Sasso Laboratory for zenith angles up to 60◦ as measured by LVD  (data points
with error bars) and generated with MUSIC (dashed curve) \cite{Music}}
\end{center}
\end{figure}

- MUSIC \cite{Music} (MUon SImulation Code) is a  simulation tool for 3 dimensional simulations of the  
muon propagation through rock. It takes into account energy losses of muons by pair production, 
inelastic scattering, brems-strahlung and ionisation as well as the angular deflection by multiple 
scattering. The program uses the standard CERN library routines and random number generators. The validity of the code has been tested comparing the simulation's results with experimental data of the azimuthal distribution of single muon intensities in the underground Gran
Sasso Laboratory for zenith angles up to 60$^{0}$ as measured by  LVD experiment \cite{LVD}. The results  \cite{Music} show a good agreement between the data and the simulations (see Fig. \ref{fig:LVD}).

- GEANT 3.21 \cite{Geant}, the detector simulation package from CERN has been used to simulate the 
interaction of the muons with the detector and for a proper calibration of the signal. Since the muon capture was not included in GEANT 3.21,  a modified version of this code developed in context of charge ratio studies  \cite{Vulpescu, Brancus_1} has been used for this study. Since both hadronic interaction models implemented in GEANT 3.21 (GEISHA \cite{Gheisha} and FLUKA \cite{Fluka}) give reasonable results for muon simulation, for this study, the default model: GEISHA has been used.

The simulation code CORSIKA has been originally designed for the
four dimensional simulation of extensive air showers with primary
energies around $10^{15}$\,eV. The particle
transport includes the particle ranges defined by the life time of
the particle and its cross-section with air. The density profile of
the atmosphere is handled as continuous function, thus not sampled
in layers of 
constant density. 
Ionization losses, multiple
scattering and the deflection in the local magnetic field
are considered. The decay of particles is simulated in exact
kinematics and the muon polarization is taken into account.
In contrast to other air shower simulations tools, CORSIKA offers
alternatively six different models for the description of the high energy
hadronic interaction and three different models for the description
of the low energy hadronic interaction. 
The threshold between the high and low energy models is set to E$_{Lab}$ = 80 GeV/n. 

\begin{figure}
\begin{center}
\includegraphics[scale=0.7]{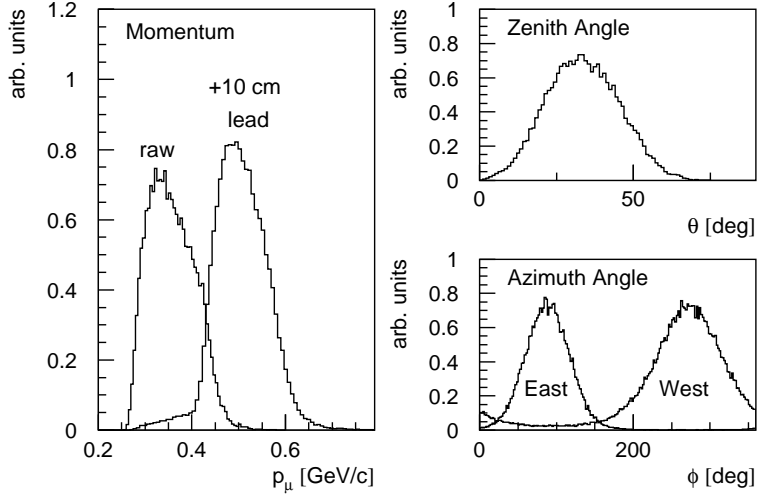}
\caption{\label{willi_det} The energetic (left) and angular acceptance (right) of the WILLI detector}
\end{center}
\end{figure}

\begin{figure}
\begin{center}
\includegraphics[scale=0.4]{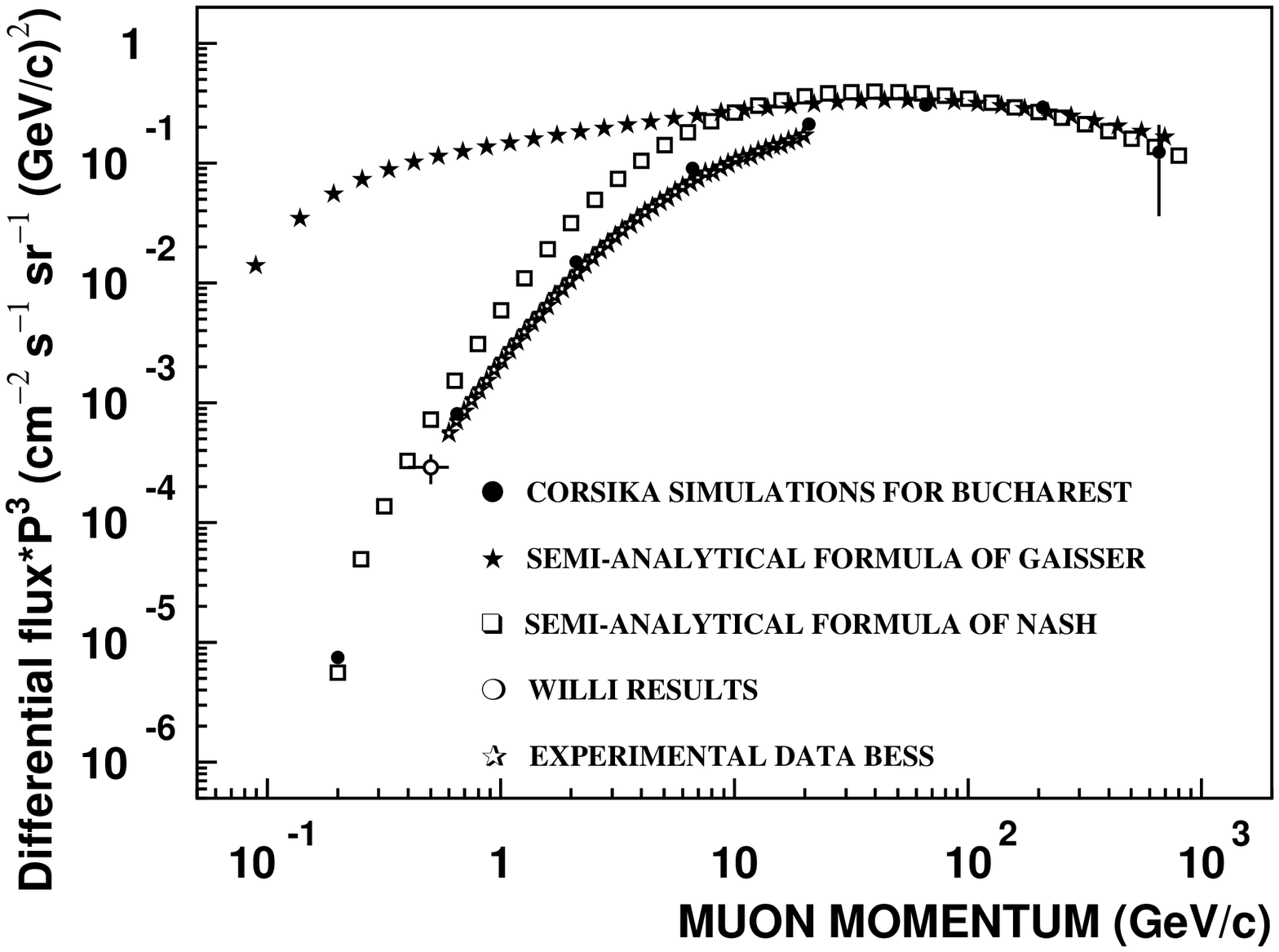}
\includegraphics[height=6.5cm, width=5.5cm]{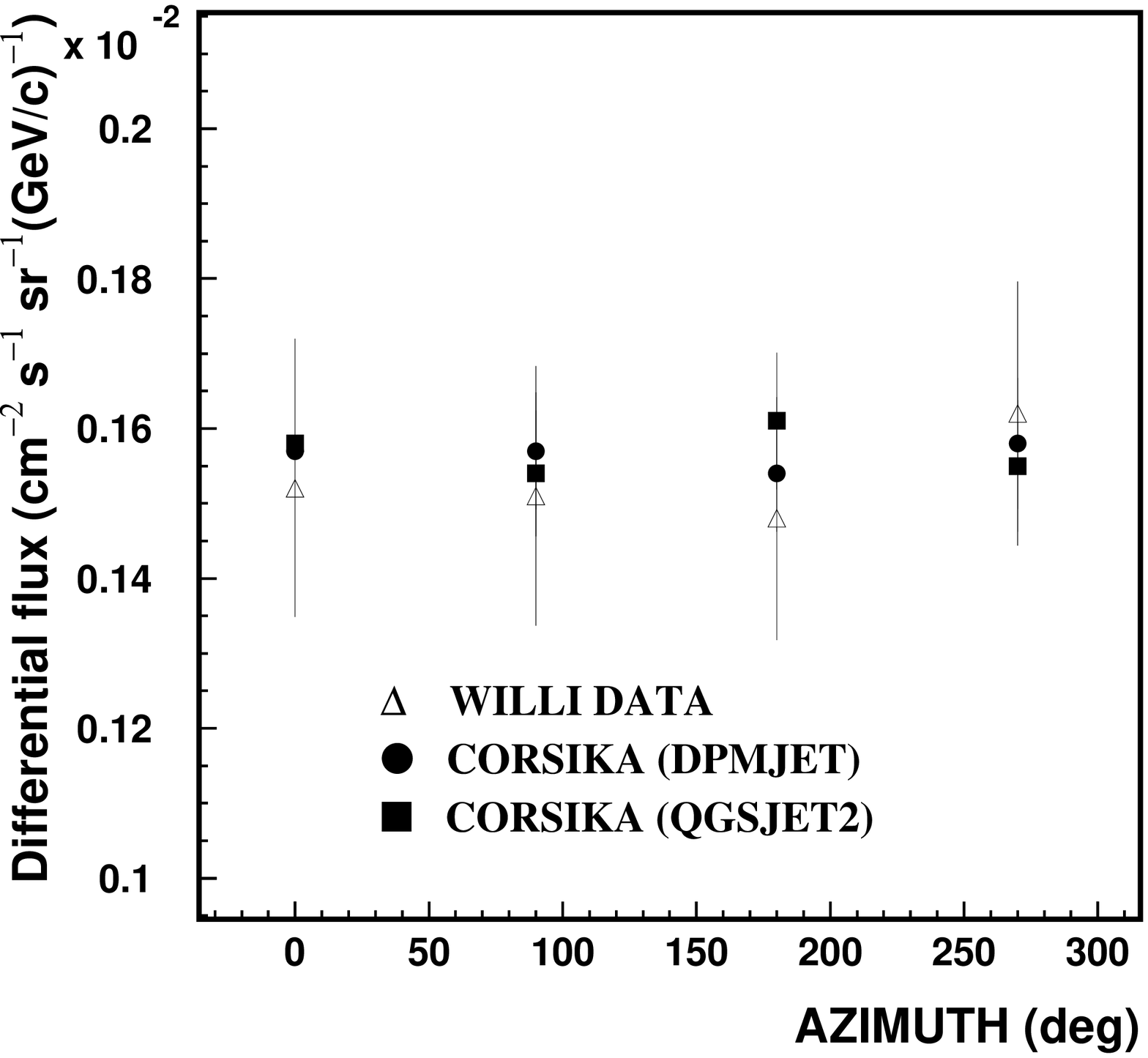}
\caption{\label{fig:muon_flux} Left: Compilation of the muon flux at surface: simulations, semi-analytical approaches and measurements; Right: Azimuthal variation of the muon flux measured by WILLI \cite{Vulpescu} and compared with CORSIKA simulations (2 different hadronic interaction models). The error bars are statistical and systematic ones.}
\end{center}
\end{figure}

The influence of the hadronic interaction models, implemented in CORSIKA, on the muon flux has been investigated in \cite{wentz1}, and all of them 
are able to reproduce the experimental muon flux data, with some differences regarding the charge ratio of muons.
From this reason, CORSIKA results given by DPMJET\cite{dpmjet} model where compared with the new  QGSJET2\cite{qgsjet2} ones  and
 with experimental findings obtained by   
WILLI  detector \cite{Mitrica}. 
The WILLI calorimeter (see Fig.~\ref{willi_det}-left), installed in IFIN-HH Bucharest \cite{Vulpescu}, is operated since
several years for measuring charge ratio and flux of atmospheric muons at energies smaller than 1 GeV (see Fig.~\ref{willi_det}-left), particularly exploring its directional dependence. The results obtained by WILLI are affected by statistical and systematic errors, the last ones due to the large geometric acceptance  of the detector (see Fig.~\ref{willi_det}-right). The experimental data agree with the simulation results (see Fig.~\ref{fig:muon_flux}) in the limit of the statistical and systematic errors. 
For higher energies, the simulations have been compared with  the experimental data of BESS detector \cite{bess}.  Fig.~\ref{fig:muon_flux}-left compares the results obtained by semi-analytical approximation with those obtained with Monte-Carlo calculations and with WILLI and BESS data. The muon flux on surface can also be estimated by semi-analytical formulae of Nash \cite{Nash} and Gaisser \cite{Gaisser}.
The validity of formula given by Gaisser  \cite{Gaisser} is restricted to muon energies $>$ 10 GeV due the fact that the muon's free decay is not considered. The formulation of Judge and Nash \cite{Nash} tuned to experimental results, may be useful for the consideration of low energy muon flux.

 In the present work, the DPMJET \cite{dpmjet}   model was used to describe the hadronic interaction 
at high energy and, the UrQMD \cite{urqmd} model for low energy processes. 
Based on the fact that the Monte-Carlo simulation code CORSIKA reproduces very well the experimental result 
(at low energies) its output  is used as input for the underground simulation. There is some sensitivity of the underground muon flux predictions to the hadronic interaction models used in the simulations.

\begin{figure}
\begin{center}
\includegraphics[scale=0.3]{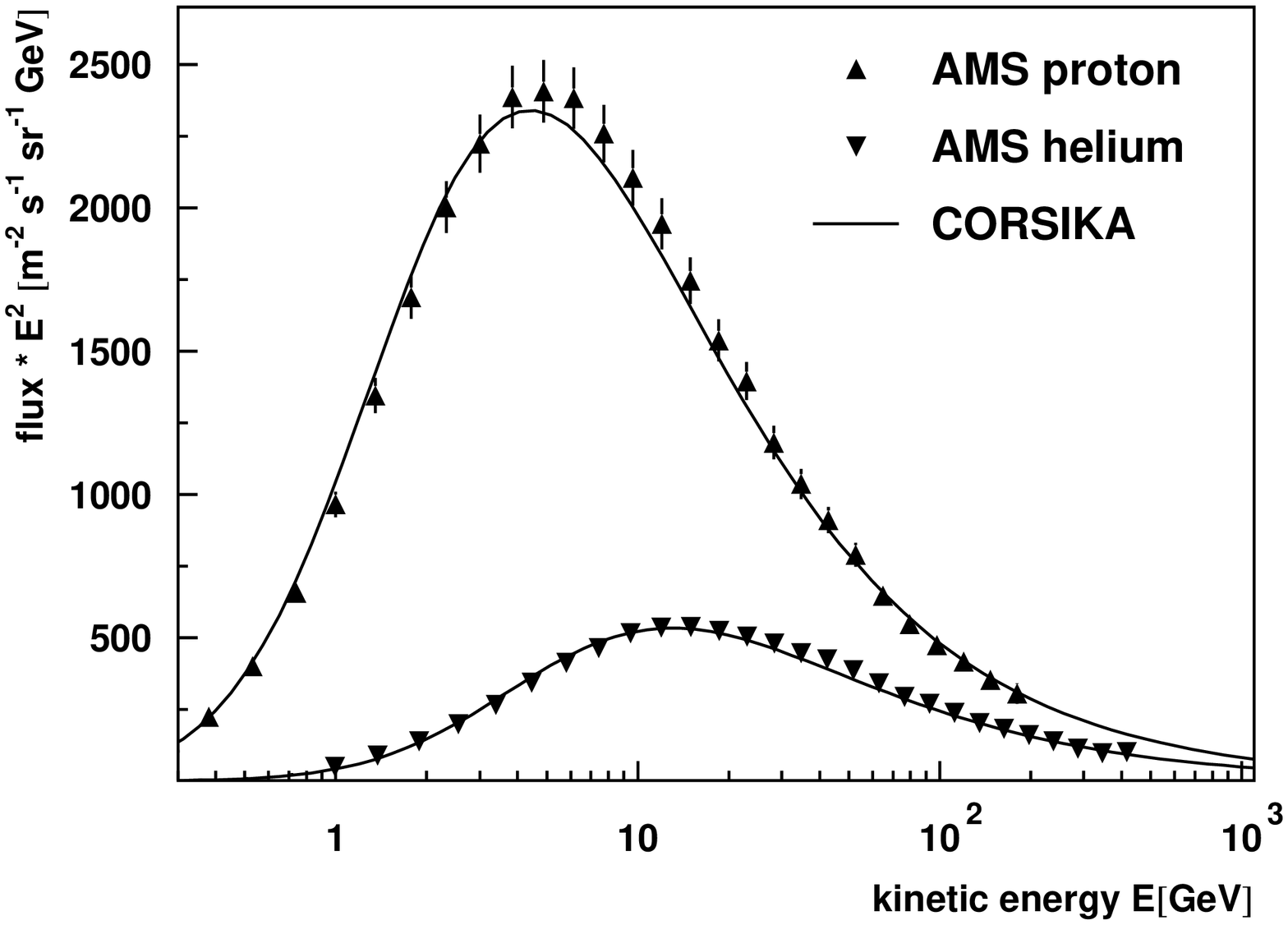}
\includegraphics[scale=0.3]{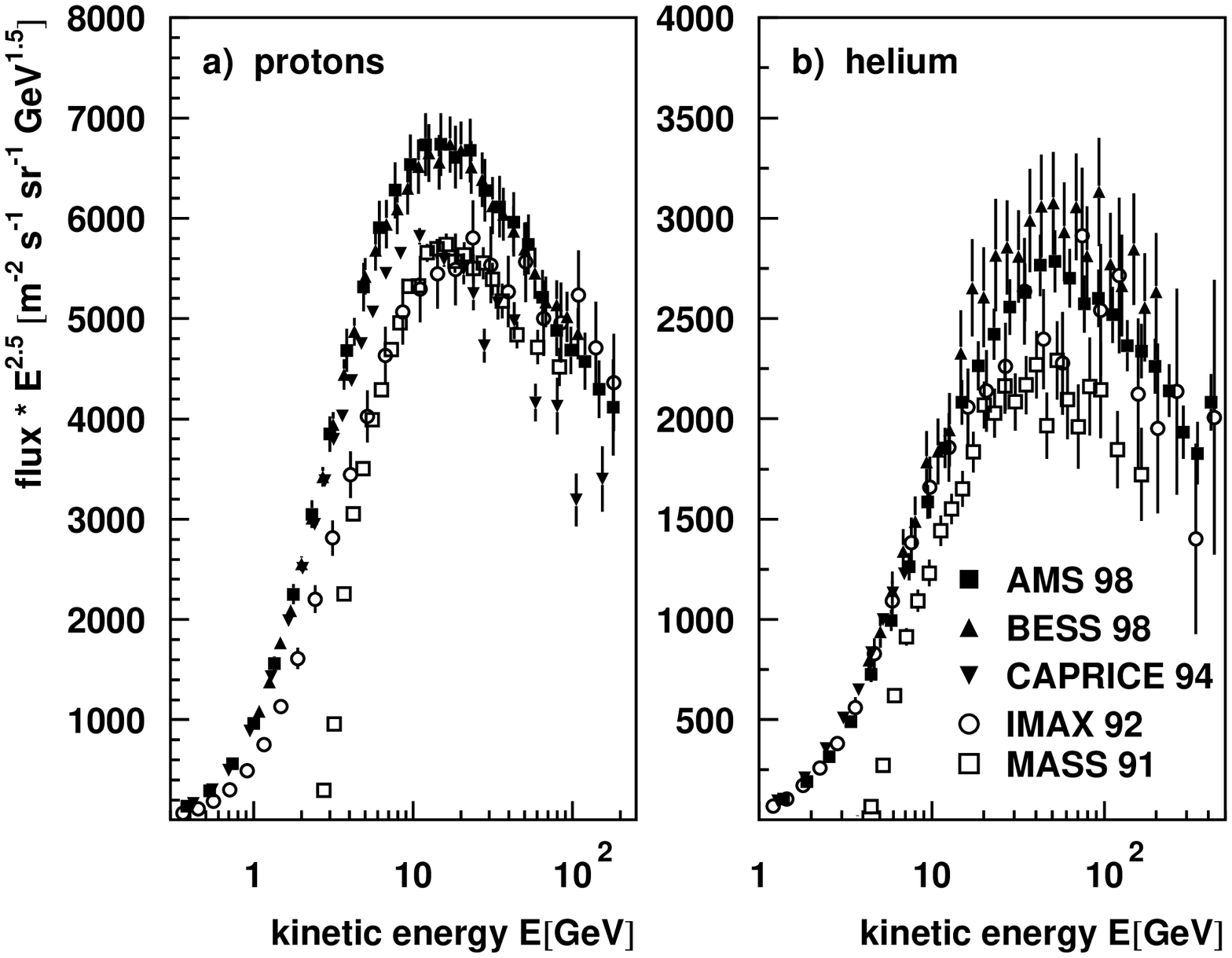}
\caption{\label{fig:primary} The primary cosmic rays spectra for protons and helium, used as input of the simulations (left). The fluxes of primary protons (center) and helium nuclei (right), measured in different balloon and satellite experiments.}
\end{center}
\end{figure}

The primary particle fluxes were introduced in the simulation's set-up using the results observed by the 
AMS prototype during a space shuttle mission \cite{ams}, using the spectrum and the ratio of  
proton and helium nuclei simulated in separate runs.  
Fig. \ref{fig:primary} shows the fluxes of primary protons (center) and helium nuclei (right) measured by different balloon and satellite borne experiments \cite{MASS, BESS_sat, CAPRICE, IMAX}, that were launched in different region with various geomagnetic cut-off. In this study, the AMS data have been used, because it covers a large range of cutoff rigidities from the geomagnetic 
equator (maximum) down to vertical cutoff rigidities less than 0.2 GV. 
The simulated muon flux generated by CORSIKA, have statistical errors (due to the limited number of considerate primary particles) and some systematic errors. The main sources of systematic errors are the primary spectrum and the hadronic interaction models. Other errors, due to particle decay or particle tracking are negligible compared to the other ones.

%%%%%%%%%%%%%%

\begin{figure}
\begin{center}
\includegraphics[scale=0.5]{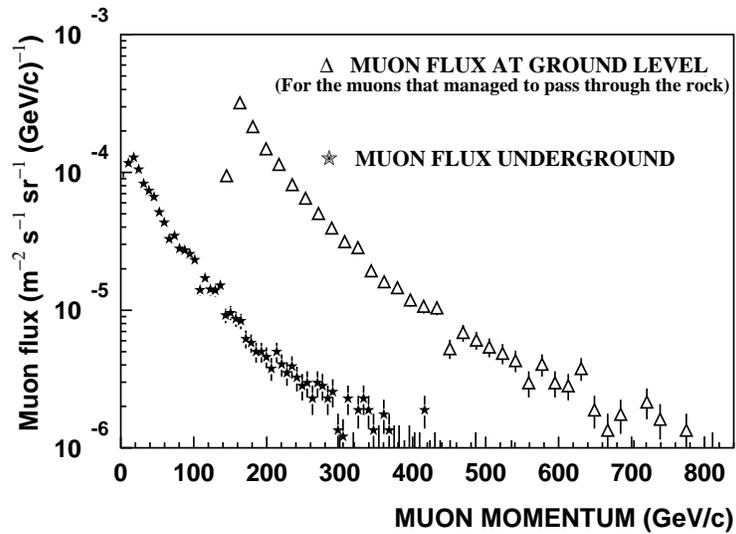}
\end{center}
\caption{\label{fig:Mine} The simulated muon flux at the surface and in the mine.}
\label{Mine}
\end{figure}

Fig.\ref{Mine} displays the muon fluxes at surface  (CORSIKA simulation) and in underground, assuming, in a first approximation, a flat overburden over the observation  level corresponding to an equivalent depth of 
600 mwe (simulated by MUSIC using the muon flux from CORSIKA as input). 
The triangles represent only the muons from surface that managed to pass through the rock. 
The energy cut off for the surviving muons is estimated to be around 150 GeV. 
By these simulation studies we estimate the expected muon rate at 600 mwe to about 
10 muons/m$^{2}$min.

\section{The apparatus}
\label{Apparatus}

\begin{figure}
\begin{center}
\includegraphics[scale=0.5]{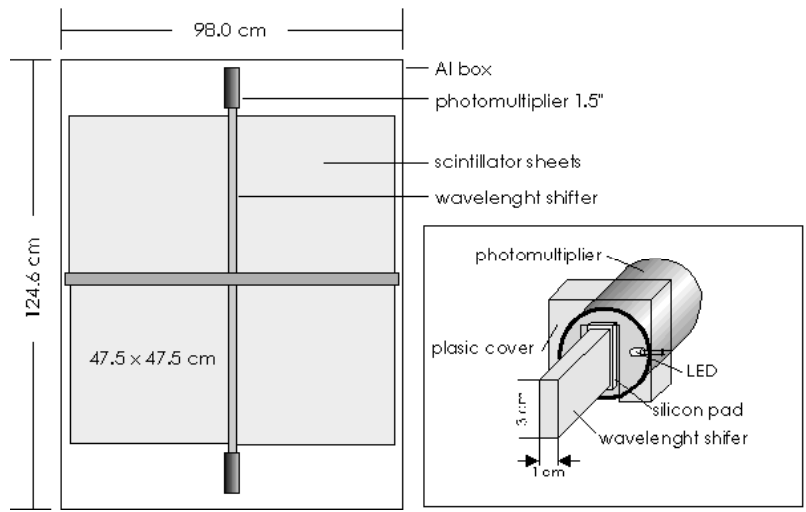}
\caption{\label{fig:module} The detection module. Design of KASCADE  \cite{Bozdog}.}
\end{center}
\end{figure}

\begin{figure}
\begin{center}
\includegraphics[scale=0.5]{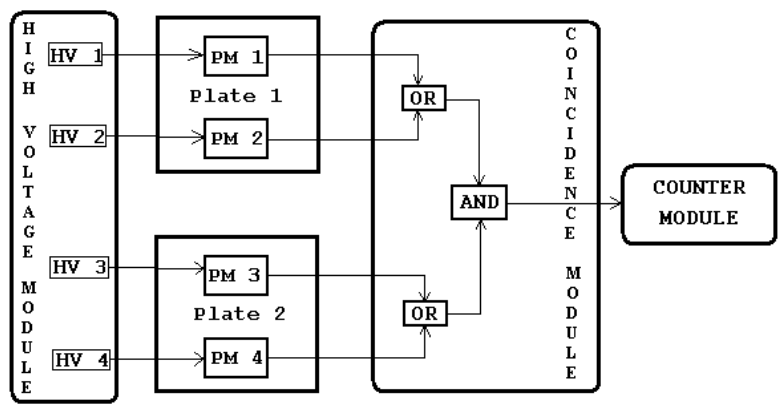}
\caption{\label{fig:electronic} The electronic detection system.}
\end{center}
\end{figure}

The mobile detector was set-up in IFIN-HH and it consists of 2 detection modules.
Each module is a scintillator plate (NE 114 type) of 0.9025 m$^2$ and 3 cm thickness, see Fig.  \ref{fig:module}, 
divided in 4 parts (0.475 x 0.475 m$^2$) \cite{Bozdog, kascade_1}, readout by two 
photomultiplier tubes (EMI 9902 type) which receives the signal trough a wave length shifter (NE 174 A type). 
The modules are arranged one on the top of the other (at 8.5 cm distance), 
in order to identify the transversing muons as coincidence event.
The signals from the 4 photomultiplier tubes  are two by two OR-ed (1 or 2) and (3 or 4) and then are putted in coincidence  (see Fig. \ref{fig:electronic}) using a gate of 50 ns so no correction due to the dead time of the detector is necessary.
A counter module registers the coincidence events.  

\begin{figure}
\begin{center}
\includegraphics[scale=0.3]{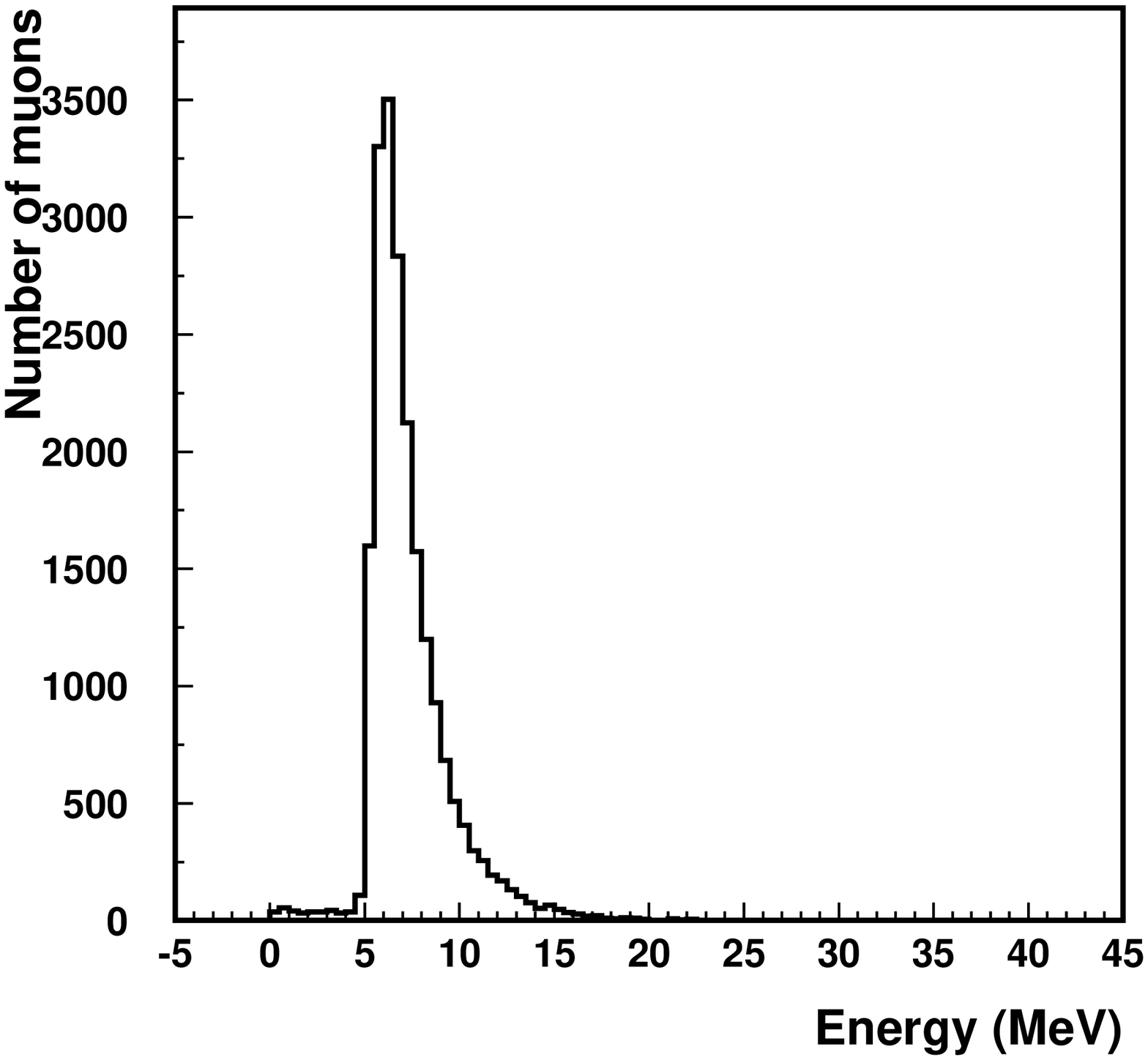}
\includegraphics[scale=0.3]{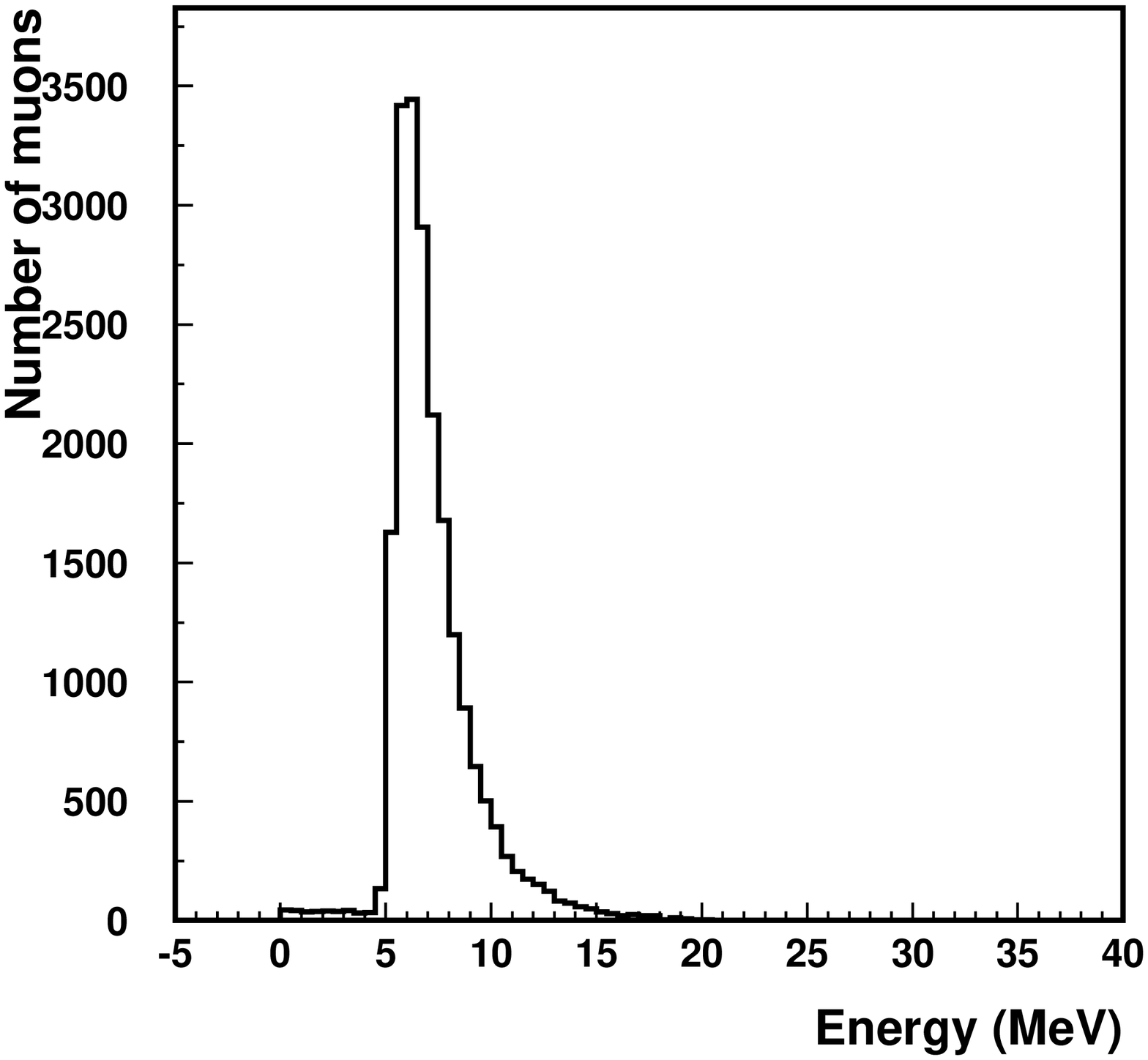}
\caption{\label{fig:edep} The energy deposit in the scintillator plates (left - the top layer, right - the bottom layer).}
\end{center}
\end{figure}

\begin{figure}
\begin{center}
\includegraphics[scale=0.7]{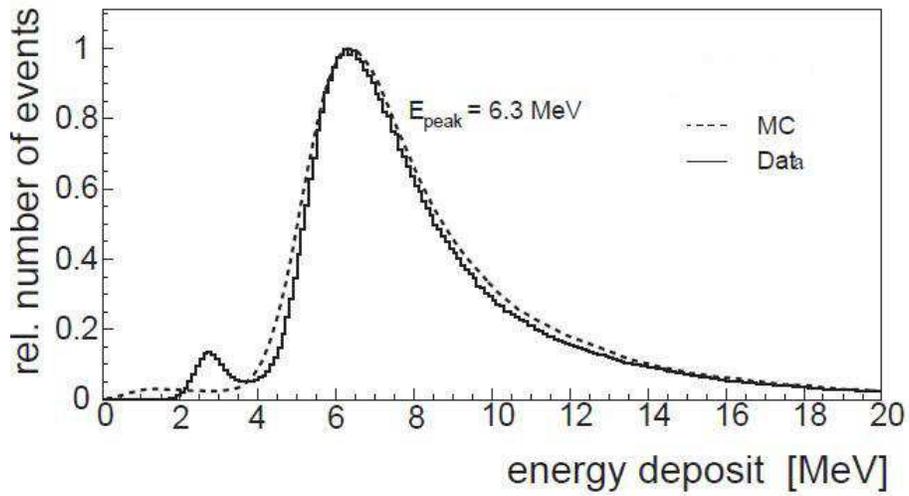}
\caption{\label{calibration} The energy calibration of one scintillator plate \cite{kascade_1}.}
\end{center}
\end{figure}

The detector response is simulated by use of the GEANT 3.21 code. 
The interaction of the muon with the active detector material and the deposit of the energy 
in the scintillator plates are analysed. 
Fig. \ref{fig:edep} displays the energy deposit by muons (generated from CORSIKA output) in each detection plate.  The signal threshold is set to 1/3 of the most probable energy deposit (6.3 MeV), i.e. to 2.1 MeV \cite{kascade_1}.
The energy calibration has been performed by minimising the normalised spectra of the expected energy deposit and the anode signal (see Fig.\ref{calibration}).

\begin{figure}
\begin{center}
\includegraphics[scale=0.8]{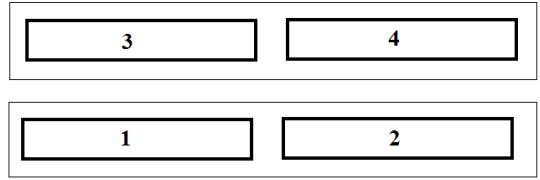}
\caption{\label{fig:disposal} The disposal of the scintillators }
\end{center}
\end{figure}

Preliminary tests had been performed on surface in order to check the counting rate for each individual scintillator and the rate of coincidences between two scintillators. The disposal of the scintillators is illustrated in Fig.\ref{fig:disposal}
 The results are displayed in Tabs.\ref{rate1} and \ref{rate2}. The rate of singles is relatively high compared with the coincidences, due to the  radiation background and electronic noise.

\begin{table}[!h]
\begin{center}
\caption {\label{rate1} The   counting rates for each scintillator plate }
\vspace*{0.5cm}
\begin{tabular}{|c|c|c|c|c|}
\hline
{\small Scintillator plate} & 1 & 2 & 3 & 4 \\
\hline
{\small Counting rate} & 515 & 550 & 541 & 508 \\
\hline
\end{tabular} 
\end{center}
\end{table}

\begin{table}[!h]
\begin{center}
\caption {The \label{rate2}  rate of coincidences between planes }
\vspace*{0.5cm}
\begin{tabular}{|c|c|c|c|c|c|c|}
\hline
{\small Scintillator plates in coincidence} & 1-2  & 3-4 & 1-3 & 2-4 & 1-4 & 2-3 \\
\hline
{\small Counting rate} & 0.56 & 0.69 & 61.54 & 60.82 & 8.75 & 9.10 \\
\hline
\end{tabular} 

\end{center}
\end{table}

%%%%%%%%%%%%%

\begin{figure}
\begin{center}
\includegraphics[scale=0.6]{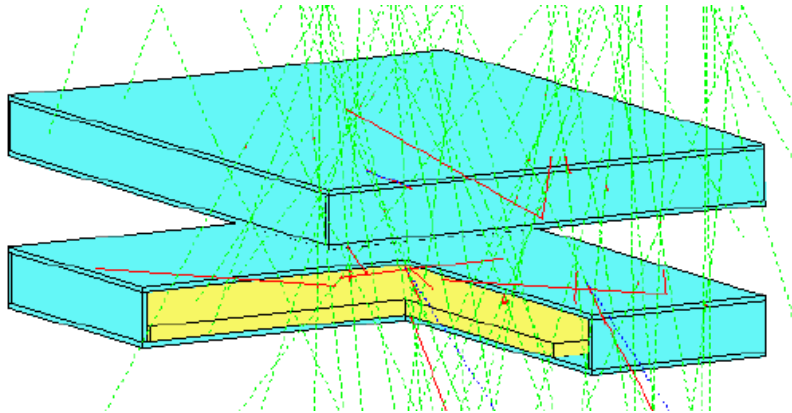}
\caption{\label{fig:geant} The schematically view of the muons interacting with the detection system, as simulated 
with GEANT code (green lines are muons, red lines are secondary electrons).}
\end{center}
\end{figure}

The  acceptance of the detector was also investigated considering the muon flux given by CORSIKA, 
taking in to account  the fact that not all muons interact with the first layer, but manage to pass to the second one 
(see Fig. \ref{fig:geant}). Thus, based on GEANT simulations, which include the signal threshold,  a correction factor of +9\% has to be applied 
on the observed muon rate.  The muon flux is given by:

\begin{eqnarray}
\phi_{\mu} = \alpha_{1} \cdot \alpha_{2} \cdot R 
\end{eqnarray}
where:  $\alpha_{1}$ = 1.09 - the correction factor given by simulation, and $\alpha_{2}$ = 1.11 - the correction factor due to the detector's surface and R - the counting rate.
 
\begin{figure}
\begin{center}
\includegraphics[scale=0.3]{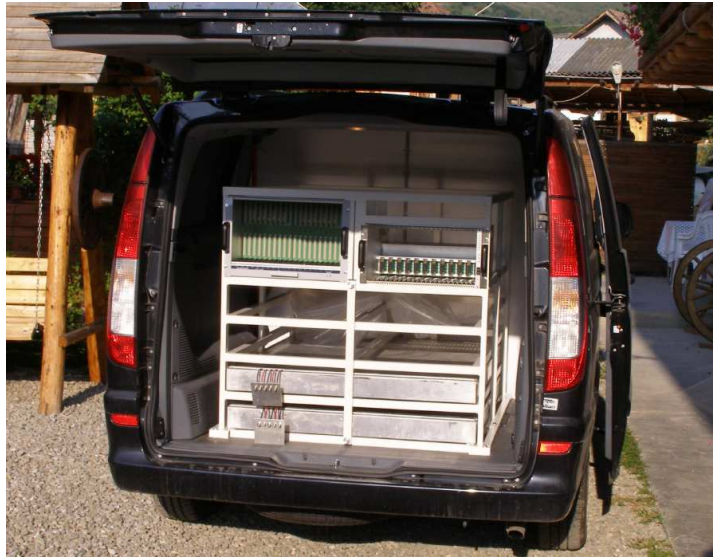}
\caption{\label{fig:car} Photo of the mobile detector mounted in a van.}
\end{center}
\end{figure}

The detector is installed on a van (see Fig. \ref{fig:car}) allowing to move quickly the system. 
The electric power for the entire system is supplied by a mobile electric generator of 1 kW power at 
230 V AC or by a 12-230 V inverter of 1 KW power which transform 12 V CC from the car's battery to 230 V AC.

Using this mobile detector, measurements of the muon flux have been performed on different altitude levels 
and geographical locations at surface and on different mines of the salt ore of the Slanic site.

\section{Measurements and results}
\label{Measurements}

The measurements of the muon flux in the underground have been performed at the Slanic site at 
3 different locations: in Unirea salt mine (in IFIN-HH lab - see Fig\ref{fig:unirea}) at 208 m below the entrance 
and in the active mine Cantacuzino, in 2 different levels, first one at 188 m and the second one at 210 m below the entrance.
All runs were performed at approx. the same hour of the day (noon) in order to reduce the influence of the solar 
activity and atmospheric conditions. 
The acquisition time for each data set was one hour. 
In Cantacuzino mine, where an access tunnel is available, the measurements have been performed using the detector 
installed on the van. In Unirea mine, the detection modules were removed from the car and transported by an elevator 
to the observation level. 
The results of the three measurement campaigns are displayed in Tab. \ref{tab:willires}.

\begin{table}[!h]
\begin{center}
\caption {\label{tab:willires} The muon flux data obtained in underground measurements. The errors are purely statistics.}
\vspace*{0.5cm}
\begin{tabular}{|c|c|c|c|}
\hline
{\small Location} & {\small Depth}  & {\small Muon flux}  & {\small mwe depth} \\
   &  (from surface) & ($m^{-2} s^{-1}$) & \\
\hline
{\small Unirea mine }                & - 208 m & 0.18 $\pm$ 0.01 &  610 $\pm$ 11\\
{\small Cantacuzino mine - Level 8}  & - 188 m & 0.19 $\pm$ 0.02 &  601 $\pm$ 21 \\
{\small Cantacuzino mine - Level 12} & - 210 m & 0.09 $\pm$ 0.01 &  790 $\pm$ 29\\

\hline

\end{tabular} 
\end{center}
\end{table}

The variation of the muon flux as a function of the water equivalent depth is given by \cite{Formaggio}:

\begin{eqnarray}
\phi_{\mu}(X) = A\cdot(X_{0}/X)^{\eta}\cdot exp(-X/X_{0})
\end{eqnarray} 
where: A = 0.03, X$_{o}$ = 1470 m.w.e. and $\eta$ = 2.5.

The difference in the muon flux measured at approximately identical physical depths in Cantacuzino (-210 m) and in Unirea 
(-208 m) is associated to the different thickness of salt rock above the detection place. 
Unirea mine is consisting of a huge cavity up to 57 m between the floor and the roof. 
In contrast, the Cantacuzino mine has a relative homogeneous rock massive above. The differences between the muon fluxes in Cantacuzino and Unirea, are caused also by the fact that the overburden of the mine is not flat, due to the irregular hill's shape above the entrance. The salt ore has a relatively regular shape (see Section \ref{LAGUNA}) and a homogeneous  composition consisting of salt (NaCl $>$ 98\%) and different impurities ($<$ 2\%) \cite{Cristache}. Based on that, the m.w.e. has been estimated using only the muon flux values. The errors reported in Tab.\ref{tab:willires} are purely statistics.

\begin{table}[!h]
\begin{center}
\caption {\label{tab:altitude} Measured muon flux at different elevations and locations 
(the altitude was determined with a GPS system). The errors are statistics, including some systematic errors due to meteorological effects. }
\begin{tabular}{|c|c|c|c|}
\hline

\small
Latitude (deg) & Longitude (deg) & Altitude (m a.s.l.)  & muon flux ($m^{-2} s^{-1}$)\\

\hline

45.29 & 25.94 & 655 $\pm$ 5 &  146,74 $\pm$ 8,24 \\
45.28 & 25.97 & 588 $\pm$ 5 &  145,30 $\pm$ 8,14 \\
45.24 & 25.94 & 408 $\pm$ 5 &  143,24 $\pm$ 7,97 \\
44.32 & 28.19 &	70  $\pm$ 5 &  128,05 $\pm$ 7,19 \\
44.40 & 26.10 &	64  $\pm$ 5 &  122,28 $\pm$ 6,76 \\
44.36 & 28.05 &	7   $\pm$ 5 &  119,07 $\pm$ 6,69 \\

\hline

\end{tabular} 
\end{center}
\end{table}

\begin{figure}
\begin{center}
\includegraphics[scale=0.4]{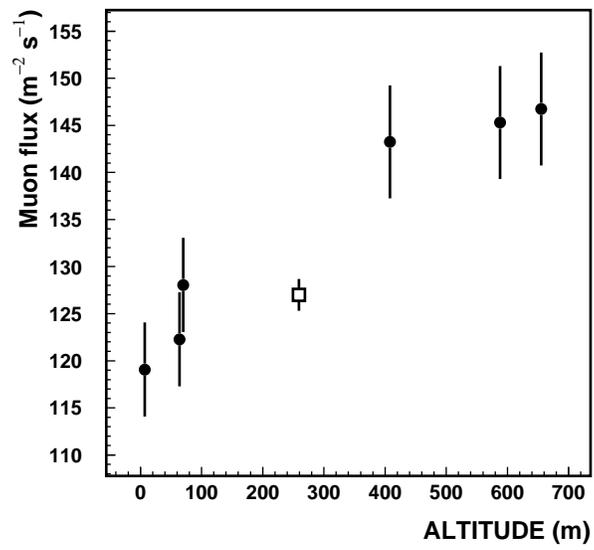}
\caption{\label{fig:altitude} Measured results of the muon flux variation with altitude in m a.s.l. The rectangle point represents the results from \cite{Greisen}. The error bars are the statistical and systematic errors.}
\end{center}
\end{figure}

\begin{figure}
\begin{center}
\includegraphics[scale=0.7]{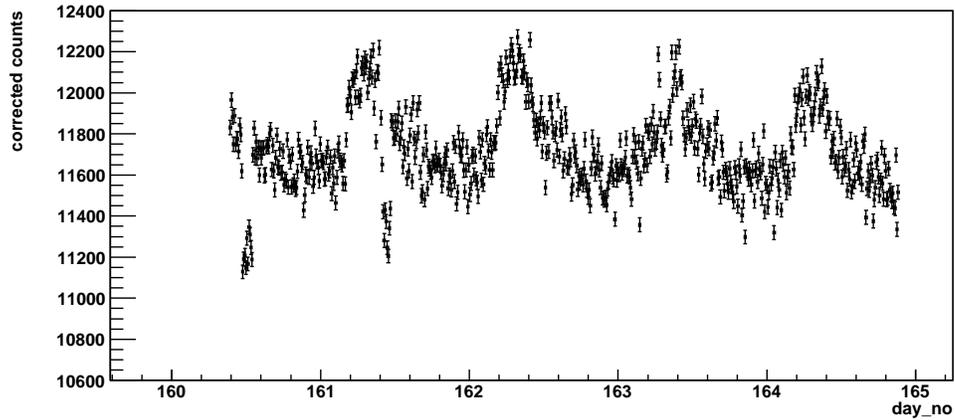}
\caption{\label{willi_june} Muon count rate as a function of day number of the year 2009 \cite{Saftoiu} divided
into 10 minutes intervals, for E $>$ 0.4 GeV. Error bars represent statistical
errors.}
\end{center}
\end{figure}

Taking advantage of the mobility of the system, the measurements of the cosmic muon flux have been performed for 
many locations on surface at different geographical positions and different elevations, from  sea level up to 655 m. 
The results are in good agreement with measurements reported previously in ref. \cite{Greisen}, that estimate a flux of 127 
muon/m$^{2}$s at 259 m a.s.l. (above sea level). 
The results are compiled in Tab.\ref{tab:altitude} and displayed in Fig. \ref{fig:altitude}. 
During the campaigns at altitudes 70 m  and 408 m the observation conditions were different (wind and low temperature) compared  to
the others, which led to different values and larger error bars.  The meteorological conditions could affect the low energy muons based on the variation of the air's density. A systematic error of aprox. 3 \% has been estimated based on WILLI measurements \cite{Saftoiu} performed over a full month (see Fig.\ref{willi_june}).
The underground muon flux cannot be affected by weather conditions due to the fact that only high energetic muons could achieve the underground observation level. The effect of the air temperature from the mine, which is constant all over the year, on the muon flux is negligible.
 The influence of the  meteorological conditions on the muon flux is presently under investigation using a portable weather station.

%\newpage
\section{Conclusions}
\label{Conclusions}

\begin{figure}
\begin{center}
\includegraphics[scale=0.4]{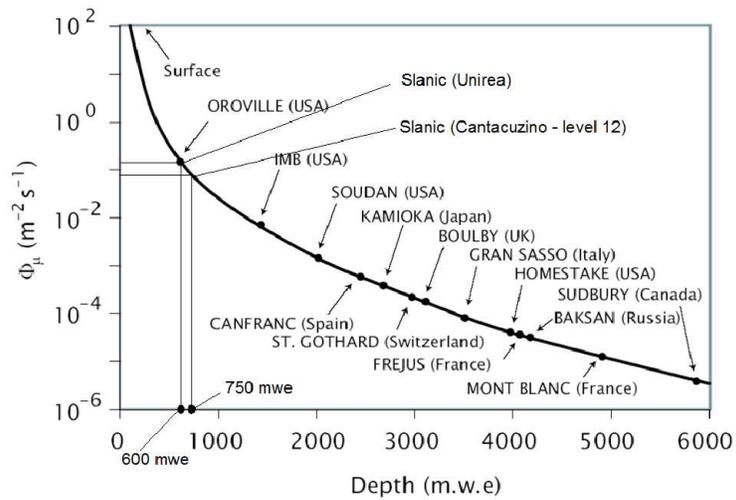}
\caption{\label{fig:depth}Muon flux as function of
MWE depths for different underground sites.}
\end{center}
\end{figure}

Suggested by  the muon flux measurements reported for other sites \cite{Carmona}, the water equivalent depth of 
different places of the  Slanic underground site were determined. 
The water equivalent depths of the Slanic mine are 610 m.w.e. for Unirea mine, 601 for Cantacuzino mine (level 8) 
and 790 m.w.e. for Cantacuzino mine (level 12), respectively.
The Slanic site is a feasible location for the GLACIER detector to be located in Unirea mine, with respect to the 
determined depth of 600 mwe (see Fig. \ref{fig:depth}). 
In addition, further promising locations for LAGUNA at the Slanic site are under consideration. 
A new cavern, 100 m below the Cantacuzino mine (see Fig. \ref{fig:slanic}) could be excavated in a reasonable time scale. 
In this case a depth of about 1000 mwe would be at disposal for experiments.

In near future, further measurements at different locations in Unirea mine will be performed, in order to get an 
improved overview on the variation of the Unirea mine's water equivalent depth. 
We expect that the muon flux varies for different locations of the mine due to the variation of the overburden 
at the Unirea mine.

Such  muon flux measurements could be also used for geological studies, 
e.g. to explore  variations in the rock density above the observation level.
The mobility of the detector implies a considerable practical flexibility of using the procedure of 
measuring muon flux differences  for various aspects. 

\section{Acknowledgements}

The present work has been possible due to the support of the Romanian Authority for Scientific Research  
by the projects {\bf PN II Capacitati 145 - CEXFIZNUC}, {\bf PN II Parteneriate 82-104/2008 - DETCOS} and {\bf PN 09 37 01 05}.
We thank the KASCADE-Grande group of the Karlsruhe Institute of Technology . for essential contributions.

\newpage

%%%%%%%%%%%%%%%%%

\end{document}